**Original Article**

# A new approach to position reconstruction in TOFPET

**Nagendra Nath Mondal**

Saha Institute of Nuclear Physics, Bidhannagar, Kolkata, West Bengal, India

**ABSTRACT**

Monte Carlo Simulation (MCS) is a state-of-the-art technique in designing sophisticated apparatus for various applications in science and technology. We adopted MCS based on GEANT (GEometry ANd Tracking) in order to design a simple time-of-flight positron emission tomography (TOFPET). In MCS studies, a new method of position reconstruction of positron-electron annihilation points has been developed so far. Simulation results show that this technique may not be useful for small animal imaging camera but might be practicable in diagnostic TOFPET camera. Specific issue of this simulation technique is discussed.

**Key words:** Monte Carlo, PET, radionuclide, scintillator, time-of-flight

Monte Carlo Simulation (MCS) is an analytical method to imitate real-life system, especially when other analyses are too mathematically complex or too difficult to reproduce. An extensive MCS based on GEANT[1] has been pursued for designing a simple PET detectors system. GEANT (acronym formed from 'GEometry ANd Tracking') is a system of detector description and simulation tools that help to describe the passage of elementary particles through matter, developed by CERN (GEANT 3.21 released in 1994) mainly for high-energy physics experiments. Now it is considerably used in other areas such as medical and biological sciences, radioprotection and astronautics. Popular codes have been written using Monte Carlo methods in GEANT for various simulation purposes. Two types of MCS codes can be utilized for simulating Single Photon Emission Computed Tomography (SPECT) and Positron Emission Tomography (PET): (i) general purpose code, which simulates particle transportation and was developed for high-energy physics or dosimetry and (ii) dedicated codes designed specifically for SPECT or PET.[2] In this study, both codes are written by users in relevant subroutines, which pass control to three phases of the run: (i) initialization, (ii) event processing and (iii) termination, wherein user of each of the three phases can implement own code in the appropriate routines. Simulated data are stored in the HBOOK/Ntuple files for analysis by the physics analysis tool PAW,[3] which is compatible graphics software for GEANT 3.21; and FORTRAN is the programming language for both the cases. The extended version of PAW is ROOT,[4] and that has been turning to GEANT 4.[5]

Information of energy and time-of-flight (TOF) of positron-electron ($e^+e^-$) annihilation γ-rays is recorded, and positions of annihilation are reconstructed from these two parameters without any conventional PET image reconstruction technique.[6] This is a new approach so far in the position reconstruction of annihilation points. Basics of physical phenomena of a TOFPET, a future plan of simple experiment, position reconstruction of $e^+e^-$ annihilations, determination of conversion factors, finally the utility of these factors will be discussed in the subsequent sections.

### Physical phenomena of TOFPET

Positron emission tomography is a method for resolving biological and physiological processes *in vivo* in a quantitative way by using radio pharmaceuticals labeled with $e^+$ emitting nuclides (short-life radioisotope) such as $^{11}C$, $^{13}N$, $^{15}O$ and $^{18}F$; and by means of the annihilation radiation using a coincidence detection technique.[6] A tagged radionuclide is injected into the patient body as labels on tracer molecules designed to probe physiological process. Positrons are emitted from the source accumulated in the affected organs, and they annihilate with electrons of the surrounding tissue after thermalization. An annihilation event mostly results into 2γ-rays almost of the same energy (511 keV) and 180° apart from each other, depending on the $e^+e^-$ angular momentum coupling. An extension of PET is a TOFPET, in which the time difference between arrivals of coincidence γ-rays is measured. In ordinary PET system, this information is ignored. Incorporating TOF information gives more weight

**Address for correspondence:**
Dr. Nagendra Nath Mondal,
Nuclear and Atomic Physics Division, Saha Institute of Nuclear Physics, 1/AF Bidhannagar, Kolkata - 700 064, West Bengal, India. E-mail: nagendra.mondal@saha.ac.in

to the more probable locations of e$^+$e$^-$ annihilation. After data acquisition, positions of the annihilation distributions are reconstructed by using the image reconstruction method (e.g., artificial neural network,[7] filtered back-projection reconstruction methods,[6] etc.).

In PET, image is always blurred due to poor spatial resolution of the system and energy resolutions of the scintillator detectors. Better image can be obtained by reducing the statistical uncertainties of TOF and energy of γ-rays. The best-quality scintillator must be chosen in order to reduce uncertainties in both the cases. Inorganic scintillator has many advantages over organic or gas-filled detectors, because of greater γ-ray stopping power due to their higher density and higher atomic Z. Scintillators with enhanced properties, such as increased X-ray conversion efficiency, faster luminescence decay times, lower afterglow and greater stability in the radiation field, are required in diagnostic imaging.

Extensive MCS studies based on GEANT have been performed in order to choose better scintillator for a TOFPET camera. For these purposes, two scintillation detectors (e.g., BaF$_2$) of the same size (30φ mm × 30 mm) are placed (face-to-face) 50 mm far from a point source of 511 keV γ-rays. In each case of the scintillator, the same number of e$^+$e$^-$ events is uniformly generated. From the energy spectra number of coincidence events (sum of two γ-ray energies, range 900-1150 keV) selected and detection efficiency are calculated accordingly. Simulation results and properties of scintillators are tabulated in the following Table 1.

The simulation results indicate that LSO has one of the best detection efficiencies among the scintillators, because of its highest density. Although it has the highest photon detection efficiency (see column 5), the overall energy resolution of LSO is not as good as NaI(Tl). Another disadvantage for general applications of this scintillator is that $^{176}$Lu is itself radioactive.[6] There remain some chances of accidental coincidence in TOFPET measurement system. If photon yield of scintillators is being considered, a BriLanCe 380 (LaBr$_3$:Ce) is one of the best choices, because of its higher density and light output than those of BaF$_2$ and faster decay time than that of any other scintillator except BaF$_2$. It has no radiation effect like LSO. Advantage of BaF$_2$ is that it has the shortest decay-time components (0.6, 0.8 ns), resulting in time resolution up to 120 ps[8]; on the other hand, a timing resolution of 850 ps at fwhm with respect to a small LSO crystal is obtained,[9] so that timing information in TOFPET system can be observed precisely if ultraviolet wavelength from BaF$_2$ is selected carefully. We have achieved 320 ps timing and 108 keV energy resolutions of BaF$_2$ scintillator (tapered type) by coupling to a Philips XP2020Q photo multiplier tube (quartz window) in the positron lifetime spectroscopy measurement.[10] Although detection efficiency of BaF$_2$ is lower than that of LSO, GSO and BGO, it can be considered for its lowest decay time and better time resolution than any other verified scintillators.

### TOFPET, a thought experiment

For systematic understanding of the TOF effect in PET system, a simple experiment is being planned; and that is shown in Fig. 1. This spectrum is obtained by MCS based on GEANT, where real parameters of scintillators, size, detector's geometry and materials are taken into account. It is planned before going into detailed study of a small animal by injecting with a short-lived e$^+$ emitting radionuclide. Our strategy is to simulate a simple PET detectors system first and reconstruct image utilizing a new technique, which is discussed in the position reconstruction section.

There are three common designs of scintillation crystal-based PET detectors: (i) continuous, (ii) block and (iii) discrete crystal detector.[2] In this simulation, the last-category detector is considered and detectors' array is chosen considering to a small animal PET camera. A simple BGO detectors-based PET system is demonstrated by K. Sonnabend et al.[11] They illustrated problems of two-dimensional imaging tomography using one-dimensional projections without considering TOF effect in their image processing. In this system, two pairs of BaF$_2$ scintillation detectors are placed face-to-face on the ±x- and the ±y-axis respectively. The size of the scintillator is 30φ mm × 30 mm, covered with a thin Al (1 mm) shield and a thick Pb (35 mm) shield in order to reflect scintillation light and suppress the accidental coincidence and the Compton scattered γ-rays respectively. The distance between the center and the surface of each scintillator is 100 mm. For a point source in the geometrical center, an expression of resolution degradation $\Delta x$ can be obtained from $\Delta x = Rt/3\sqrt{2}d$ where R: radius of the detector, $t$: thickness of the detector and $d$: detector-to-detector distance.[12] Providing all the input parameters of the detectors system, $\Delta x = 0.53$ mm is achieved, which is comparable with the simulation results. The detector-to-detector distance is good enough to observe the difference of times between the coincident γ-rays that are traveling with the velocity of light. The face-to-face distance of two detectors is 200 mm. Due to Pb collimator, the effective distance of the present system is reduced to 140 mm. Diameter of the small-size rat or a mice may not exceed this size; hence a small animal can easily pass through these detectors' array. For experimental research purpose, this configuration of detectors is quite good. Generated number of e⁺e⁻ events is $10^6$, resulting in a source strength ~0.27 µCi (3-4 orders lower than the injected source strength of a 70-kg human body in PET scan); and coincidence counting rate is ~2.

**Table 1: Properties of different scintillators with simulated results**

| Scintillators | Density (g/cm$^3$)* | Light yield* (Photons/keV), η | Detection efficiency, ε (%) | η × ε |
|---|---|---|---|---|
| 1. NaI(Tl) | 3.67 | 38 | 0.252 | 9.58 |
| 2. BaF$_2$ | 4.88 | 1.8 | 0.416 | 0.75 |
| 3. BriLanCe 380 | 5.29 | 63 | 0.304 | 19.15 |
| 4. YAP(Ce) | 5.55 | 18 | 0.084 | 1.51 |
| 5. GSO | 6.71 | 15 | 0.996 | 14.94 |
| 6. BGO | 7.13 | 9 | 1.77 | 15.93 |
| 7. LSO | 7.40 | 32 | 1.34 | 42.88 |

*www.detectors.saint-gobain.com

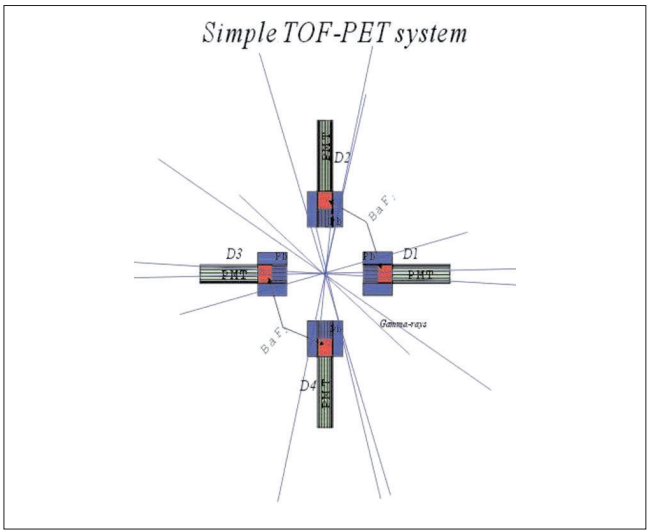

**Figure 1:** Simple PET detectors system. e⁺e⁻ annihilation photon lines are shown clearly

The advantage of TOFPET over PET is to increase the Signal-to-Noise (S/N) ratio, the ability to handle high counts and much lower accidental coincidence rates than those of PET. In the conventional PET system, an event is valid only when two detectors are fired by coincident 511 keV γ-rays within a timing window typically 8-14 ns. Those events are recorded along the lines of response (LOR) by the face-to-face detectors, and there is no way to separate the Compton scattered events from the real events which are on the LOR in the image reconstruction algorithm.

In TOFPET the actual time difference in the arrival of annihilated 2γ-rays at the face-to-face detectors is recorded. Recently for a few scintillators, time resolutions of a few hundred picoseconds are achievable. This can be used to constrain the reconstruction algorithm as weight and helps to localize the annihilation site to within a few centimeters. The approximate improvement in S/N ratio over that obtained with non-TOFPET is given by[13]

$$\left(\frac{S}{N}\right)_{TOF} \approx \sqrt{\frac{2D}{C\Delta t}} \times \left(\frac{S}{N}\right)_{non-TOF} \quad (1)$$

where D is the diameter of the object being imaged, C is the velocity of light, $\Delta t$ is the timing resolution of the system. From Eq. (1), it can be seen that S/N can be improved by improving the time resolution of the scintillator - i.e., if $\Delta t$ reduces to 10 times, the S/N ratio improves by ~3 times.

A spatial resolution [SR(fwhm) = 2.35 × σ, where σ is the uncertainty of Gaussian distribution] plays an important role in PET system for producing real image of the diagnostic object. SR refers to the area on the ground and the minimum distance between two adjacent features that an imaging system can distinguish. Current trend is to achieve SR less than 1 cm. Therefore, instead of individual detectors, most PET camera makers are utilizing pixels, which are coupled with position-sensitive photomultiplier tubes. Spatial resolutions both in the x- and the z-axis are determined. Equal number of e⁺e⁻ annihilation events are generated considering a point-source distribution at each point along ± x- and ± z-axis respectively. The numbers of 511 keV γ-rays detected by four detectors are plotted along the vertical axis and corresponding generation points along the horizontal scale. Simulated points are fitted with Gaussian function. From the fitted data, values of SR = 28.12 ± 12.43 cm and 8.62 ± 0.08 cm respectively for the ± x- and the ± z-axis are obtained. Similarly, better values of SR = 2.18 ± 0.26 cm and 1.24 ± 0.01 cm for the ± x- and ± z-axis respectively are obtained when only coincidence events are taken into account. From these two analyses, it is concluded that SR along the z-axis is better than that along the x-axis because of the detectors' arrangement and γ-ray shielding. It is concluded from these results that coincidence technique is very important in image reconstruction for improving the SR and S/N ratio. It can be argued also that the resolving power of PET is better than that of SPECT.

### *Position reconstruction - a new approach*

A two-dimensional imaging technique is considered for reconstructing e⁺e⁻ annihilation positions. In the MCS data, both the energy and the time information of detected γ-rays by individual BaF$_2$ detectors are recorded. Generalized equations for reconstructed positions of e⁺e⁻ annihilation points along the x- and the y-axis respectively are given below:

$$X = \sum_{i=1}^{n/2}[T_i |\cos(i-1)\theta|$$
$$-T_{i+n/2}|\cos(i+n/2-1)\theta|]\times C/2 \quad (2)$$

$$Y = \sum_{i=1}^{n/2}[T_i |\sin(i-1)\theta|$$
$$-T_{i+n/2}|\sin(i+n/2-1)\theta|]\times C/2 \quad (3)$$

where $T_i$ and $T_{i+n/2}$ are the TOF of annihilated photons detected respectively by $D_i$ and $D_{i+n/2}$ detectors, $n$ is the even number of detectors in a ring and $\theta$ ($=2\pi/n$) is the angle between the neighboring detectors. In this particular case [Figure 1], n = 4 and $\theta$ = 90° and the corresponding reconstructed position along the x-axis is

$$X_{1,3} = (T_1 - T_3)\times C/2 \quad (4)$$

where $X_{1,3}$ is the reconstructed position, $T_1$ and $T_3$ are the values of TOF of annihilation γ-rays detected respectively by D1 and D3 detectors along the x-axis. Similarly, $Y_{2,4}$ can be reconstructed using Eq. 3 and by $T_2$ and $T_4$ data. From the energy-time correlation data, those events are selected, which are completely stopped by the scintillator and make a full energy deposit of 511 keV. Considering the distance between the event generation positions and the detector, a time window (0.3-1.4 ns) and a threshold of energy conservation (900-1150 keV) are applied. It is noticed that the reconstructed positions (obtained by Eq. 4) shift significantly except position (0,0,0) and do not satisfy the ideal situation, which poses a big question. It is observed that TOF distribution is not purely Gaussian; there is a long tail at the right-hand side of the mean position of the TOF spectrum. As a result, $T_1$-$T_3$ spectrum contains two long tails on both sides. This tail appears because of decay-times (slow and fast) components of $BaF_2$ and multiple reflections of scattered photons inside the scintillator.[14,15] In this simulation, no discrimination of decay times is applied. Therefore, it is essential to find some position conversion factors for reproducing the original event-generation positions.

### *Determination of conversion factors*

Positron-electron annihilation events are generated at different positions beginning from (0,0,0) up to (6,0,0) cm and considering the point-source distribution for determining the conversion factors. Selecting the energy-time correlation events from the data as before, each annihilation position is reconstructed by Eq. (4); and those positions are plotted along the vertical axis with corresponding real position at the horizontal scale. Data points are fitted with least square fitting method, and the slope obtained is m = 1.12 ± 0.03; and the intercept, c = 0.15 ± 0.09 cm. It can be seen from c and m values that the fitted line is not passing through (0,0,0).

However, introducing $c$ and $m$, Eq. (4) can be written as

$$X^c_{1,3} = (X_{1,3} - c)/(2m) \quad (5)$$

where $X^c_{1,3}$ is the corrected and reconstructed position of $X_{1,3}$. A factor '2' appears in Eq. (5) due to minimization of the SR of the two face-to-face detectors. Similarly, reconstructed $Y_{2,4}$ position can be corrected. Eq. (5) reflects the ideal situation. Conversion factors are used in the image reconstruction process, because the full spectra of TOFs that are not digitized are taken into account. A digitized form of TOF spectrum is possible to obtain when scintillator size of a detector is very small, like pixel at least 10 times smaller than the $BaF_2$ size, currently used in this simulation; and detector-to-detector distance is very large ~100 cm (real PET system). In that case, conversion factors may not be useful in image reconstruction, the discussion of which is beyond the scope of the present article.

### *Utility of conversion factors*

In order to understand the importance of conversion factors, $e^+e^-$ annihilation events are generated at two arbitrary positions (e.g., x, y, z = –6,0,0 and 6,0,0 cm). The size of the original distribution of $e^+e^-$ events is about 0.52 $cm^3$, which is taken into account because it resembles a small-size tumor.

Using Eq. (5), positions were reconstructed; and corresponding spectrum is shown in Figure 2A. The data is fitted with two Gaussians and is convoluted using the fitting parameters after normalizing with the real spectra. Real and convoluted data are plotted on the same spectrum, and that is shown in Fig. 2B. The dark area (crossed by two Gaussians) attributes the spatial resolution along the x-axis consistent with the simulated value mentioned earlier, and no distinction could be possible between two or more tumor positions that are lying inside this area. The coincidence detection efficiency in this measurement is ~0.06%.

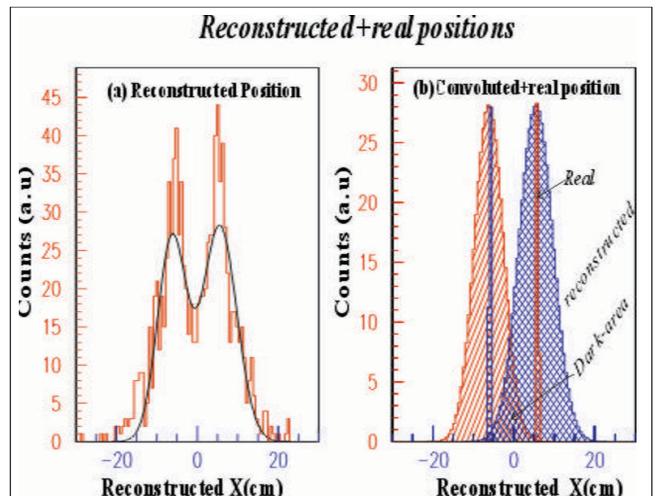

**Figure 2:** (A) Reconstructed and (B) convoluted position distributions

No doubt, SR of this system is extremely larger than that of real PET system[15] or small-animal imaging camera, where SR is about a few mm. It is suggested that reconstructed spatial resolution should be <1 mm in all directions as opposed to the ~10 mm reconstructed image resolution typical in human whole-body studies.[2] However, this newly developed method might be applicable for image reconstruction in TOFPET system without conventional PET image reconstruction technique, which saves huge computational time of iteration. One very important solution to reduce the SR is to select the decay constant 0.6 ns by choosing the UV light of $BaF_2$ scintillator by digital oscilloscope,[8] investigation of which is in progress by considering the real PET detectors' geometry.

Considering other image-reconstruction methods, Filtered Back Projection (FBP) is one of the conventional and widely used techniques in PET. Although it has computational simplicity, FBP fails to consider important effects like noise, attenuation, scatter and blur. In order to incorporate all these effects, it is necessary to include additional parameters in the image-reconstruction algorithms, as a result, it increases the image-processing time and memory of the hard disk. Different numbers of iterations are required that easily lead to noise amplification with image quality deterioration.[2] Presented advanced technique is completely free from the iteration procedures; hence it will reduce the amount of source injected into the patient body, the data-taking time (for a few minutes), computational image-processing time (less than hours) and save memory by reducing the size of the data (within a few Mb).

## Conclusion

A new method has been established so far for TOFPET $e^+e^-$ annihilation position reconstruction in two dimensions. By storing the data of TOF and the energy of annihilation γ-rays, the positions are reconstructed. Position-conversion factors are important parameters, and those have been achieved. Instead of setting window of time typically at 8-14 ns (corresponding difference of path length 180 cm is quite large in comparison to the diameter of a PET ring) at present TOFPET experiment, individual TOF data recording of each detector seems to be practical for image reconstruction by this technique. MCS of a real TOFPET system is in progress, where 48 small-size detectors in a single ring are considered. Design of a simple TOFPET system and a study of biological sample *in vivo* are advancing.

## Acknowledgments

I would like to thank Prof. S. Bhattacharya and Prof. B. N. Ganguly for their initiative with respect to this project. The author is grateful to Prof. Bhattacharya for the critical reviewing of the manuscript.

**Source of Support:** Prof. S. Bhattacharya and Prof. B. N. Ganguly, **Conflict of Interest:** None declared.